\magnification=1200

\font\bigbf=cmbx10  scaled\magstep2 \vskip 0.2in
\centerline{\bigbf Noncommutative quantum mechanics } \vskip
0.1in \centerline{\bigbf and the Aharonov-Casher effect  }

\vskip 0.4in \font\bigtenrm=cmr10 scaled\magstep1
\centerline{\bigtenrm  B. Mirza$^{\dag \ddag}$ and M.
Zarei$^{\dag}$} \vskip 0.2in

\centerline{\sl $^{\dag}$Department of Physics, Isfahan University of Technology, Isfahan 84154, Iran }
\vskip 0.1in
\centerline{\sl $ ^{\ddag}$Institute for Studies in Theoretical Physics and Mathematics, }
\centerline{\sl P.O.Box 5746, Tehran, 19395, Iran}
\vskip 0.1in

\centerline{\sl E-mail: b.mirza@cc.iut.ac.ir}

\vskip 0.2in \centerline{\bf ABSTRACT} \vskip 0.1in

In this work a new method is developed to investigate the
Aharonov-Casher effect in a noncommutative space. It is shown
that the holonomy receives non-trivial kinematical corrections.



\vfill\eject

\vskip 1in

\centerline{I. \bf \ Introduction } \vskip 0.1in

In the last few years, theories in noncommutative space have been
studied extensively (for a review see [1]). Noncommutative field
theories are related to M-theory compactification [2], string
theory in nontrivial backgrounds [3] and quantum Hall effect [4].
Inclusion of noncommutativity in quantum field theory can be
achieved in two different ways: via Moyal $\star$-product on the
space of ordinary functions, or defining the field theory on a
coordinate operator space which is intrinsically
noncommutative[1,5]. The equivalence between the two approaches
has been nicely described in [6]. A simple insight on the role of
noncommutativity in field theory can be obtained by studying the
one particle sector, which prompted an interest in the study of
noncommutative quantum mechanics [7,8,9,10,11,12,13,14]. In these
studies some attention was paid to the Aharonov-Bohm effect [15].
If the noncommutative effects are important at very high
energies, then one could posit a decoupling theorem that produces
the standard quantum field theory as an effective field theory
and that does not remind the noncommutative effects. However the
experience from atomic and molecular physics strongly suggests
that the decoupling is never complete and that the high energy
effects appear in the effective action as topological remnants.
Along these lines, the Aharonov-Bohm effect has already been
investigated in a noncommutative space [16].  In this work, we
will develop a new method to obtain the corrections to the
topological phase of the Aharonov-Casher effect, where we know
that in a commutative space the line spectrum does not depend on
the relativistic nature of the dipoles. The article is organized
as follows; in section 2, we discuss the Aharonov-Casher effect
on a commutative space. In section 3, the Aharonov-Casher effect
in a noncommutative space is studied and a generalized formula for
holonomy is given.

\vskip 0.2in \centerline{II. \bf \ The Aharonov-Casher effect }
\vskip 0.1in

In 1984 Aharonov and Casher (AC) [17] pointed out that the wave
function of a neutral particle with nonzero magnetic moment $\mu
$ develops a topological phase when traveling in a closed path
which encircles an infinitely long filament carrying a uniform
charge density. The AC phase has been measured experimentally
[18]. This phenomenon is similar to the Aharonov-Bohm (AB) effect.
The similarities and the differences of these two phenomena and
possible classical interpretations of the AC effect have been
discussed by several authors [19,20,21]. In Ref. [17], the
topological phase of the AC effect was derived by considering a
neutral particle with a nonzero magnetic dipole moment moving in
an electric field produced by an infinitely long filament
carrying a uniform charge density. If the particle travels over a
closed path which includes the filament, a topological phase will
result. This phase is given by

$$ \phi_{AC}= \oint (\vec \mu \times \vec E) \cdot  d{\vec r}  \eqno(1) $$

\noindent   where $\vec \mu =\mu \vec \sigma $ is the magnetic
dipole moment and $\vec \sigma = (\sigma_1,\sigma_2,\sigma_3)$,
where $ \sigma_i \ (i=1,2,3)$ are the $2 \times 2$ Pauli matrices.
It is possible to arrange that the particle moves in the x-y plane
and travels over a closed path which includes an infinite filament
along z-axis. The electric field in the point $ \vec r = x \hat i
+ y \hat j $, where $\hat i$ and $ \hat j $ are unit vectors in
the direction of the positive x and y axes, is given as

$$ \vec E = {\lambda \over {2 \pi (x^2 + y^2 )}}(x \hat i
+ y \hat j)  \eqno(2) $$

 \noindent where $\lambda $ is the linear charge density of the
filament and the phase is given by

$$  \phi_{AC}= \mu \sigma_3 \oint (\hat{k} \times \vec E ) \cdot  d
 \vec r =  \mu \sigma_3 \lambda   \eqno(3) $$

\noindent  where $\hat k $ is a unit vector along z axis. This
phase is purely quantum mechanical and has no classical
interpretation. The appearance of $\sigma_3 $ in the phase
represents the spin degrees of freedom. We see that different
components acquire phases with different signs. This is also one
of the points that distinguishes the AC effect from the AB effect
[22]. In this part, we briefly explain a method for obtaining
Eq.(3). The equation of motion for a neutral spin half particle
with a nonzero magnetic dipole moment moving in a static electric
field $\vec E $ is given by

$$ (i \gamma_{\mu} \partial^{\mu}+ {1\over 2}\mu \sigma_{\alpha \beta}F^{\alpha \beta}-m)
 \psi=0 \eqno(4) $$

\noindent or it can be written as

$$ (i \gamma_{\mu} \partial^{\mu}  -i\mu  \vec {\gamma} \cdot \vec {E} \gamma_{0} -m)  \psi=0
\eqno(5) $$

\noindent where   $\vec \gamma = (\gamma^1, \gamma^2 , \gamma^3)$
and $ \gamma-$matrices are defined by

$$ \gamma^0= \left(\matrix{I & 0\cr 0 & -I\cr }\right) \ \ \ \ \ \
 \  \   \ \ \ \ \ \ \
 \gamma^i=\left(\matrix{ 0 & \sigma_i \cr -\sigma_i & 0  \cr
 }\right)\ \ \ \ \
 \eqno(6) $$

 \noindent We define

$$ \psi = \  e^{{\bf a} f} \psi_{0} \eqno(7) $$

\noindent where { \bf a } is the matrix to be determined below, $
f$ is a time independent scalar phase, and $ \psi_0 $ is a
solution of the
 Dirac equation

$$ (i \gamma_{\mu} \partial^{\mu}-m)\psi_{0} =0 \eqno(8) $$

\noindent Writing $\psi_0 $  in terms of $\psi $ and multiplying
 (8) by $ e^{{\bf a} f} $  from the left, we obtain

$$ e^{{\bf a} f} (i \gamma^{\mu} \partial_\mu  -m )e^{-{\bf a} f}\psi = 0 \eqno(9) $$

\noindent

$$ (i  e^{{\bf a} f} \gamma^\mu e^{-{\bf a} f}\partial_\mu
- i e^{{\bf a} f} \gamma^i e^{-{\bf a}  f} {\bf a} \  \partial_i f
-m)\psi=0 \eqno(10) $$

\noindent Comparing Eq.(10) with Eq.(5), we find that {\bf a} and
$f $ must satisfy

$$ {\mu}\vec {\gamma}\cdot \vec E \gamma_0 =   (\vec {\gamma} \cdot \vec
{\nabla}f) {\bf a} \ \ \ , \ \ \ {\bf a} \gamma_\mu = \gamma_\mu
{\bf a} \eqno(11) $$

\noindent The matrix {\bf a} can be expressed by some linear
combination of the complete set of $4 \times 4 $ matrices $ 1,
\gamma_5, \gamma_\mu, \gamma_\mu \gamma_5 $ and $\sigma_{\mu
\nu}={i\over 2}[\gamma_\mu , \gamma_\nu]$. The second member of
Eqs.(11) cannot be satisfied if all $\gamma_1, \gamma_2 $ and
$\gamma_3 $ are present in Eq.(10). However, it is possible to
satisfy it if the problem in question can be reduced to the
planar one. This indicates that the AC topological phase can arise
 only in two spatial dimensions. Therefore, let us consider the particle moving
 in $x-y$ plane in which case only the matrices $\gamma_1 $ and
 $\gamma_2 $ are present in (11), and moreover, $\partial_3 \psi $  and $ E_3 $
 vanish. The choice $-i \sigma_{12}\gamma_0 $ represents a
 consistent Ansatz.  From the first equation in
 (11), we get

$$ \vec \nabla f= \mu ( \hat k \times \vec E) \eqno(12) $$

\noindent and the phase is given by

$$ \eqalignno{\phi^{(0)} & = \sigma_{1 2} \gamma_0  \oint \vec \nabla f\cdot d\vec r  \cr
&=\mu \sigma_{1 2} \gamma_0 \oint (\hat{k} \times \vec E )\cdot d
\vec r \cr
 &=\mu \left(\matrix{\sigma_3&0\cr
                              0& -\sigma_3\cr}\right)\oint (\hat{k} \times \vec E )\cdot d
\vec r & (13) \cr} $$

\vskip 0.2in \centerline{III. \bf \ The Aharonov-Casher effect in
a noncommutative space } \vskip 0.1in

The noncommutative Moyal spaces can
 be realized as spaces where coordinate operator $ \hat x^\mu $
 satisfies the commutation relations

$$ [ \hat x^\mu , \hat x^\nu ] = i \theta^{\mu \nu } \eqno(14) $$

\noindent where $ \theta^{\mu \nu} $ is an antisymmetric tensor
and is of space dimension (length$)^2 $. We note that space-time
noncommutativity, $\theta^{0 i}\neq 0 $, may lead to some
problems with unitarity and causality. Such problems do not occur
 for the quantum mechanics on a noncommutative space with a usual
 commutative time coordinate. The noncommutative models specified
 by Eq.(14) can be realized in terms of a $\star$-product: the
 commutative algebra of functions with the usual product
 f(x)g(x) is replaced by the $\star$-product Moyal algebra:

 $$( f\star g)(x)= exp \ [{i\over 2} \theta_{\mu \nu
 }\partial_{x_\mu}\partial_{y_{\nu}}]\ f(x)g(y)|_{x=y} \eqno(15)$$

\noindent  As for the phase space, inferred from string theory,
we choose

$$ [\hat x_i , \hat x_j ]= i \theta_{i j}, \ \ \  [\hat x_i, \hat
p_j ]=i\hbar \delta_{ij}, \ \ \ [\hat p_i, \hat p_j]=0. \eqno(16)
$$

\noindent The noncommutative quantum mechanics can be defined by
[7-14],

$$ H( p, x) \star \psi( x)=E \psi( x)\eqno(17) $$

\noindent  The equation of motion for a neutral spin half
particle with a nonzero magnetic dipole moment moving in a static
electric field $\vec E $ is given by

$$ (i \gamma_{\mu} \partial^{\mu}+ {1\over 2}\mu \sigma_{\alpha \beta}F^{\alpha \beta}-m)\star \psi=0
\eqno(18) $$

\noindent or it can be written as

$$ (i \gamma_{\mu} \partial^{\mu}  -i\mu  \vec {\gamma} \cdot \vec {E} \gamma_{0} -m)\star \psi=0
\eqno(19) $$

\noindent We define

$$ \psi =   e^{{\bf a} f} \psi_{0}  \eqno(20) $$

\noindent  where  {\bf a} is the matrix already defined ($\bf a
\gamma_{\mu}=\gamma_{\mu}\bf a $), $f$ is a time independent
scalar phase, and $ \psi_0 $ is a solution of the
 Dirac equation

$$ (i \gamma_{\mu} \partial^{\mu}-m)\psi_{0} =0 \eqno(21) $$

\noindent and Eq.(19) can be written as

$$  (i \gamma_{\mu} \partial^{\mu} e^{{\bf a} f}) \psi_{0} - (i \mu  \vec {\gamma} \cdot \vec {E}
\gamma_{0})\star ( \ e^{{\bf a} f}  \psi_{0}) =0 \eqno(22) $$

\noindent After expanding the second term in Eq.(22) up to the
first order of the noncommutativity parameter $ \theta_{ij}=
\theta \epsilon_{ij} $  and defining $k_j $ as

$$ \partial_{j} \psi_{0} = (i k_j ) \psi_{0}  \eqno(23) $$

\noindent the final result up to first order in $\theta $  is
given by

$$ [i (\gamma^i \partial_i {\bf a} f )-i \mu \vec{\gamma}\cdot \vec{E}
\gamma_0 -{i\over 2}   \theta^{ij}[\partial_i (i \mu
\vec{\gamma}\cdot \vec{E} \gamma_0)](i k_j)-{i\over 2} \theta^{i
j} [\partial_i (i \mu \vec{\gamma}\cdot \vec{E} \gamma_0)]{\bf a}\
\partial_j f] \ e^{{\bf a} f}  \psi_{0}=0  \eqno(24) $$

\noindent or we get the following equation ($\bf a
\gamma_{\mu}=\gamma_{\mu}\bf a $),

$$ [i (\gamma^i \partial_i  {\bf a} f   )-i \mu \vec{\gamma}\cdot \vec{E}
\gamma_0 -{i\over 2}   \theta^{ij}[\partial_i (i \mu
\vec{\gamma}\cdot \vec{E} \gamma_0)](i k_j)-{i\over 2} \theta^{i
j} [\partial_i (i \mu \vec{\gamma}\cdot \vec{E} \gamma_0)]{\bf a}\
\partial_j f]\psi_{0} =0
\eqno(25) $$

\noindent It should be noted that expansion of $\vec E $ up to
first order in $\theta $ leads to an additive correction to the
commutative holonomy and does not cause a new non-topological
behaviour. A similar situation happens in the noncommutative
Aharonov-Bohm  effect. By expanding $ f $ up to first order in $
\theta $

$$ f= f^{(0)}+\theta f^{(1)}+ ... \eqno(26) $$

 \noindent we obtain the following equations,

$$ [ \mu \vec \gamma \cdot \vec E \gamma_0 - (\gamma^i  \ \partial_i f^{(0)})
\ {\bf a}]\psi_0=0 \eqno(27) $$

\noindent which is equivalent to Eq.(11) and

$$ [\gamma^i {\bf a} \ \partial_i f^{(1)} + {1\over 2} \mu \varepsilon^{ij }k_j \partial_i
(\mu \vec \gamma \cdot\vec E \gamma_0)- {i\over 2}
\varepsilon^{ij}\partial_i (\mu \vec \gamma \cdot\vec E
\gamma_0)  {\bf a} \   \partial_j f^{(0)}]\psi_0=0 \eqno (28)
$$

\noindent  By choosing  $ {\bf a} =-i \sigma_{12}\gamma_{0} $ and
after a straightforward calculation we get

$$\nabla f^{(0)}= \mu (\hat k \times \vec E )  \eqno(29) $$

\noindent and the phase is given by
$$ \eqalignno{\phi^{(0)} & = \sigma_{1 2} \gamma_0 \oint \nabla f^{(0)}\cdot d \vec r \cr
&=\mu \sigma_{1 2} \gamma_0 \oint (\hat k \times \vec E )\cdot d
\vec r \cr
 &=\mu \left(\matrix{\sigma_3&0\cr
                              0& -\sigma_3\cr}\right)\oint (\hat k \times \vec E)\cdot d
\vec r & (30) \cr} $$

\noindent Substituting Eq.(29) in (28) yields

$$ [i \gamma^i \sigma_{12} \gamma_0 \partial_i f^{(1)} - {1\over
2}\varepsilon^{ij}k_j \partial_i(\mu \vec \gamma \cdot \vec E
\gamma_0) + {1\over 2}\varepsilon^{ij}\partial_i(\mu \vec \gamma
\cdot \vec E \gamma_0)\sigma_{12} \gamma_0 \mu (\hat k \times
\vec E)_j]\psi_0=0 \eqno(31) $$

\noindent After a long but straightforward calculation, the
following correction to $\phi^{(0)}$ for a neutral particle with
nonzero magnetic dipole moment $ \mu $ and with spin up or down
($\mp $) is obtained

\vskip 0.3in

$$ \eqalignno{ \triangle \phi_{\theta} & = \theta \sigma_{1 2} \gamma_0  \oint \vec \nabla f^{(1)}
\cdot d\vec r \cr &= {\theta \over 2}\sigma_{1 2} \gamma_0
\varepsilon^{ij} \ ( \mu \oint  k_j (\partial_i E_2 dx_1-
\partial_i E_1 dx_2)   \cr
&  \mp \oint[(\mu \partial_i  E_2) \ \mu (\hat k \times \vec E)_j
dx_1 - (\mu \partial_i E_1) \ \mu (\hat k \times \vec E)_j dx_2] )
& (32) \cr}  $$

\noindent The first term is a velocity dependent correction and
does not have the topological properties of the commutative AC
effect and could modify the phase shift. The second term is a
correction to the vortex and does not contribute to the line
spectrum. Using  the following notation

$$   \vec k  \propto \vec v  \ \ \ \ \ ,  \  \ \  \mu (\vec k \times \vec E )
\propto \vec A^{(0)}  \eqno(33) $$

\noindent  the integral in Eq.(32) can be mapped to the
corrections which have already been obtained for the Aharonov-Bohm
effect in [16, Eq.(2.19)] where $ \vec v $ and $ \vec A^{(0)} $
are velocity of the particle and vector potential in the
Aharonov-Bohm  effect.


 \noindent The total phase shift for the AC effect is given by

$$ \phi_{total} = \phi_{0} + \triangle \phi_{\theta} \eqno(34) $$

$$ \phi_{0}= \mu \lambda  \eqno(35)$$

\noindent  where $\lambda $ is charge per unit length along the
filament. $\triangle \phi_{\theta}$ can be estimated for a
circular path and the contribution to the shift coming from
noncommutativity, relative to the usual shift of phase, is given
by

$$  {\triangle \phi_{\theta}\over { \phi_0}} \simeq  {\theta  \over {R \lambda_n} } +
{\theta \phi_0 \over {\pi R^2}}
  \eqno(36)  $$

\noindent where $ \lambda_n $ is the wavelength of the neutral
particle  and $ R $ is the radius of the approximate path. The
noncommutative contributions are very tiny. The experimental
observations on the AC phase shift [18] can be used to put a limit
on the noncommutativity parameter $\theta $. In [18], a crystal
neutron interferometer has been used and thermal neutrons travel
in half of their paths in a constant electric field. The path is
not circular but can be approximated by a circle with a radius
which is about 1 cm ($L=2.53 \ cm; R \simeq L sin(22.5^\circ) $
in [18]). Fitting Eq.(36) into the accuracy bound of the
experiment [18], we obtain

$$ {\triangle \phi_{\theta}\over { \phi_0}} \leq 25\% \eqno(37) $$

$$ \sqrt {\theta} \leq 10^7 GeV^{-1} \eqno(38) $$

\noindent The low energy (thermal) neutrons in the experimental
test of the AC effect [18] cause a higher limit for $\theta $ as
 compared to other limits recently obtained [23,24,25]; however, we hope that future experiments with high
energy neutrons leads to a better limit for the noncommutativity
parameter in the AC effect.

\noindent

\vskip 0.2in \centerline{\bf \ \ Acknwoledgements} \vskip 0.2in
Our thanks go to the Isfahan University of Technology and
Institute for Studies in Theoretical Physics and Mathematics for
their financial support.

\vskip 0.2in
\centerline{\bf \ \  References} \vskip 0.1in

\noindent [1] \ M. R. Douglas and N. A. Nekrasov, Rev. Mod. Phys.
{\bf 73} (2001) 977-1029. hep-th/0106048.

\noindent [2] \ A. Connes, M. R. Douglas and A. Schwarz, JHEP {\bf
9802} (1998) 003, hep-th/9808042.

\noindent [3] \ N. Seiberg and E. Witten, JHEP {\bf 9909} (1999)
032. hep-th/9908142.

\noindent [4] \ L. Susskind, hep-th/0101029.

\noindent [5] \ M. Chaichian, A. Demichev and P. Presnajder, Nucl.
Phys. {\bf B567} 360 (2000), hep-th/9812180.

\noindent [6] \ L. Alvarez-Gaume, S.R. Wadia, Phys. Lett. {\bf
B501} 319 (2001), hep-th/0006219.

\noindent [7] \ L. Mezincescu, hep-th/0007046.

\noindent [8] \ V. P.Nair, Phys. Lett. {\bf B505 } (2001) 249,
hep-th/0008027.

\noindent [9] \ M. Chaichian, M.M. Sheikh-Jabbari and A.Tureanu
Phys. Rev. Lett. {\bf 86}(2001) 2716, hep-th/0010175.

\noindent [10] \ J. Gamboa, M. Loewe and J. C. Rojas, Phys. Rev.
{\bf D64} (2001) 067901, hep-th/0010220.

\noindent [11] \ S. Belluchi , A. Nersessian and C. Sochichiu,
Phys. Lett. {\bf B522} (2001) 345, hep-th/0106138.

\noindent [12] \ C. Acatrinei, JHEP {\bf 0109} (2001) 007,
hep-th/0107078.

\noindent [13] \ J. Gamboa, M. Loewe, F.M. Mendez,J.C. Rojas,
Int. J. Mod. Phys. {\bf A17} (2002) 2555-2566, hep-th/0106125.

\noindent [14] \ J. Gamboa, M. Loewe, F. Mendez and J.C.Rojas,
Mod. Phys. Lett. {\bf A16} (2001) 2075-2078, hep-th/0104224.

\noindent [15] \ Y. Aharonov and D. Bohm, Phys. Rev. {\bf 115}
(1958) 485.

\noindent [16] \ M. Chaichian, A. Demichev, P. Presnajder, M.M.
Sheikh-Jabbari and A. Tureanu, Phys. Lett. {\bf B527}, 149-154
(2002), hep-th/0012175.

\noindent [17] \ Y. Aharonov and A. Casher, Phys. Rev. Lett. {\bf
53} (1984) 319.

\noindent [18] \ A. Cimmino, G.I. Opat, A.G. Klein, H.Kaiser,
S.A. Werner, M.Arif and R. Clothier, Phys. Rev. Lett. {\bf 63}
(1989) 380.

\noindent [19] \ A.S. Goldhaber, Phys. Rev. Lett. {\bf 62} (1989)
482.

\noindent [20] \ T.H. Boyer, Phys. Rev. {\bf A36} (1987) 5083.

\noindent [21] \ Y. Aharonov, P. Pearl and L. Vaidman, Phys. Rev.
{\bf A37} (1988) 4052.

\noindent [22] \ Xiao-Gang He and Bruce H.J. McKeller, Phys Lett.
{\bf B256} (1991) 250.

\noindent [23] \ I. Hinchliffe, N. Kersting, Y.L. Ma, Review of
the Phenomenology of Noncommutative Geometry, hep-ph/0205040.

\noindent [24]\ P. Schupp, J. Trampetic, J. Wess and G. Raffelt,
hep-ph/0212292.

\noindent [25] \ P. Minkowski, P. Schupp and J. Trampetic,
hep-th/0302175.

\noindent

\vfill\eject

\bye